\documentclass[twocolumn, superscriptaddress,pre,notitlepage]{revtex4-1}
\usepackage[T1]{fontenc}
\usepackage[latin9]{inputenc}
\usepackage{amsmath}
\usepackage{amssymb}
\usepackage{graphicx}

\makeatletter
\usepackage{bm}\usepackage{amsfonts}
\usepackage{color}

\makeatother

\begin{document}
\title{Geodesic path for the optimal nonequilibrium transition: Momentum-independent
protocol}
\author{Geng Li}
\affiliation{Graduate School of China Academy of Engineering Physics, Beijing 100193,
China}
\author{C. P. Sun}
\affiliation{Graduate School of China Academy of Engineering Physics, Beijing 100193,
China}
\affiliation{Beijing Computational Science Research Center, Beijing 100193, China}
\author{Hui Dong}
\email{hdong@gscaep.ac.cn}

\affiliation{Graduate School of China Academy of Engineering Physics, Beijing 100193,
China}
\begin{abstract}
Accelerating controlled thermodynamic processes requires an auxiliary
Hamiltonian to steer the system into instantaneous equilibrium states.
An extra energy cost is inevitably needed in such finite-time operation.
We recently develop a geodesic approach to minimize such energy cost
for the shortcut to isothermal process. The auxiliary control typically
contains momentum-dependent terms, which are hard to be experimentally
implemented due to the requirement of constantly monitoring the speed.
In this work, we employ a variational auxiliary control without the
momentum-dependent force to approximate the exact control. Following
the geometric approach, we obtain the optimal control protocol with
variational minimum energy cost. We demonstrate the construction of
such protocol via an example of Brownian motion with a controllable
harmonic potential.
\end{abstract}
\maketitle

\section{Introduction\label{SSec-one}}

The quest to accelerate a system evolving toward a target equilibrium
state is ubiquitous in various applications \citep{Ogbunugafor2016,Ahmad2017,Iram2020,Ilker2022,Albash2018,Takahashi2019,GueryOdelin2019,GueryOdelin2022}.
In the biological pharmacy, pathogens are expected to evolve to an
optimum state with maximal drug sensitivity \citep{Ogbunugafor2016,Ahmad2017,Iram2020,Ilker2022}.
Controlling the evolution of pathogens towards the target state with
a considerable rate is critically relevant to confronting the threat
of increasing antibiotic resistance and determining optimal therapies
for infectious disease and cancer. In adiabatic quantum computation,
the solution of the optimization problem is transformed to the ground
state of the problem Hamiltonian \citep{Albash2018,Takahashi2019,GueryOdelin2019,GueryOdelin2022}.
Speeding up the computation requires to steer the system evolving
from a trivial ground state to another nontrivial ground state within
finite time. These examples require to tune the system from one equilibrium
state to another one within finite time.

The scheme of shortcuts to isothermality was developed as such a control
strategy to maintain the system in instantaneous equilibrium states
during evolution processes \citep{Li2017,Patra2017,Dann2019}. Relevant
results have been applied in accelerating state-to-state transformations
\citep{Albay2019,Albay2020,Albay2020a,Jun2021}, raising the efficiency
of free-energy landscape reconstruction \citep{Li2019,Li2021}, designing
the nano-sized heat engine \citep{Martinez2017,Plata2020,Tu2021,Chen2022},
and steering biological evolutions \citep{Iram2020,Ilker2022}. Additional
energy cost is required due to the irreversibility in the finite-time
driving processes. Much effort has been devoted to find the minimum
energy requirement in the driving processes \citep{Schmiedl2007a,GomezMarin2008,Seifert2012,Bonanca2014,Plata2019,Deffner2020,Chen2021}.
We recently proved that the optimal path for the shortcut scheme is
equivalent to the geodesic path in the geometric space spanned by
control parameters \citep{Li2022}. Such an equivalence allows us
to find the optimal path through methods developed in geometry.

Implementing such shortcut scheme remains a challenge task since the
driving force required in the shortcut scheme is typically momentum-dependent
\citep{GueryOdelin2019,Jun2021,GueryOdelin2022}. One solution is
to use an approximate scheme \citep{Sels2017,Li2021} to replace the
exact one. Such scheme has been applied in the underdamped case to
obtain a driving force without any momentum terms. The key idea is
to use an approximate auxiliary control without the momentum-dependent
terms.

\begin{figure*}[!tp]
\includegraphics[width=17cm]{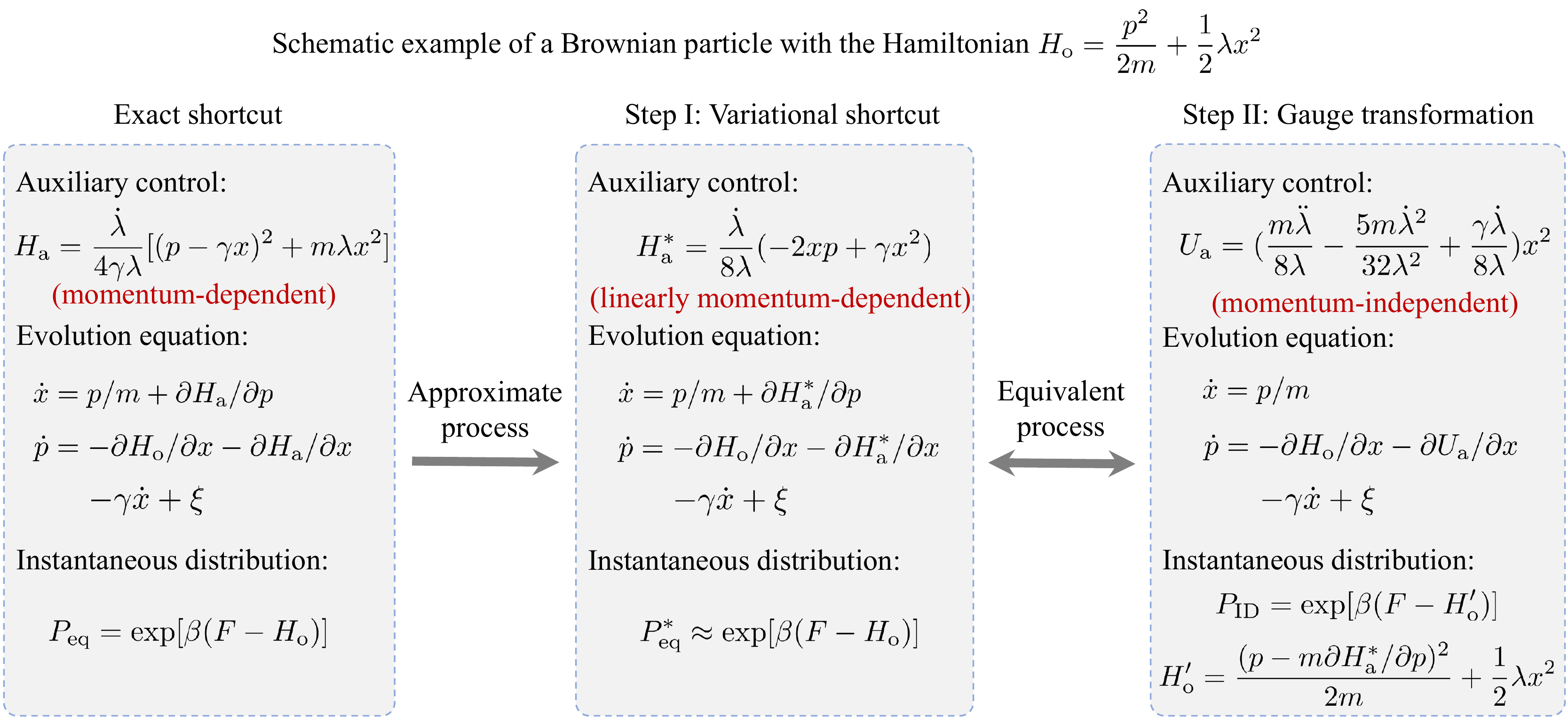} \caption{(Color online) Schematic example of a Brownian particle with the Hamiltonian
$H_{\mathrm{o}}=p^{2}/(2m)+\lambda x^{2}/2$. The dynamical evolution
of the system is governed by the Langevin equation, where $\gamma$
is the dissipation coefficient. In the shortcut scheme, a momentum-dependent
auxiliary control $H_{\mathrm{a}}$ is added to escort the system
distribution in the instantaneous equilibrium distribution $P_{\mathrm{eq}}$.
In step I, a variational shortcut scheme is used to obtain an approximate
auxiliary control $H_{\mathrm{a}}^{*}$ which can keep the system
in the approximate equilibrium distribution $P_{\mathrm{eq}}^{*}$.
In step II, a gauge transformation scheme is used to obtain a momentum-independent
auxiliary control $U_{\mathrm{a}}$ which can maintain the system
distribution in $P_{\mathrm{ID}}$.}
\label{Schematic}
\end{figure*}

In this work, we employ the variational method and the geometric approach
to find an experimental protocol with minimum dissipation for realizing
the shortcut scheme. In Sec. \ref{SSec-two}, we briefly introduce
the shortcut scheme and the geometric approach for finding the optimal
control protocol with minimum energy cost. In Sec. \ref{SSec-three},
we apply a variational method to overcome the difficulty of the momentum-dependent
terms in the driving force. As illustrated in Fig.$\ \ref{Schematic}$,
the variational method is separated into two steps. In step I, a variational
shortcut scheme is used to obtain an approximate auxiliary control
without high-order momentum-dependent terms. In step II, a gauge transformation
scheme is used to remove the linear momentum-dependent terms and an
experimentally testable protocol is obtained. In Sec. \ref{SSec-four},
we demonstrate our protocol through a Brownian particle moving in
the harmonic potential with two controllable parameters. In Sec. \ref{SSec-five},
we conclude the paper with additional discussions.

\section{Geometric approach and the auxiliary Hamiltonian\label{SSec-two}}

In this section, we briefly review our geodesic approach of the shortcut
to isothermality and show the possible experimental difficulties to
apply the obtained auxiliary Hamiltonian.

Consider a system with the Hamiltonian $H_{\mathrm{o}}(\vec{x},\vec{p},\vec{\lambda})=\sum_{i}p_{i}^{2}/(2m)+U_{\mathrm{o}}(\vec{x},\vec{\lambda})$
immersed in a thermal reservoir with a constant temperature $T$.
Here $\vec{x}\equiv(x_{1},x_{2},\cdots,x_{N})$ are coordinates, $\vec{p}\equiv(p_{1},p_{2},\cdots,p_{N})$
are momentum, $m$ is mass, and $\vec{\lambda}(t)\equiv(\lambda_{1},\lambda_{2},\cdots,\lambda_{M})$
are time-dependent control parameters. In the shortcut scheme, an
auxiliary Hamiltonian $H_{\mathrm{a}}(\vec{x},\vec{p},t)$ is added
to steer the evolution of the system along the instantaneous equilibrium
states $P_{\mathrm{eq}}=\exp[\beta(F-H_{\mathrm{o}})]$ in the finite-time
interval $t\in[0,\tau]$ with the boundary conditions $H_{\mathrm{a}}(0)=H_{\mathrm{a}}(\tau)=0.$
Here $F\equiv-\beta^{-1}\ln[\iint d\vec{x}d\vec{p}\exp(-\beta H_{\mathrm{o}})]$
is the free energy and $\beta=1/(k_{\mathrm{B}}T)$ is the inverse
temperature with the Boltzmann constant $k_{\mathrm{B}}$. The probability
distribution of the system's microstate $P(\vec{x},\vec{p},t)$ evolves
according to the Kramers equation
\begin{equation}
\frac{\partial P}{\partial t}=\sum_{i}[-\frac{\partial}{\partial x_{i}}(\frac{\partial H}{\partial p_{i}}P)+\frac{\partial}{\partial p_{i}}(\frac{\partial H}{\partial x_{i}}P+\gamma\frac{\partial H}{\partial p_{i}}P)+\frac{\gamma}{\beta}\frac{\partial^{2}P}{\partial p_{i}^{2}}],\label{eq:kramerseq}
\end{equation}
where $H\equiv H_{\mathrm{o}}+H_{\mathrm{a}}$ is the total Hamiltonian,
and $\gamma$ is the dissipation coefficient. The auxiliary Hamiltonian
is proved to have the form $H_{\mathrm{a}}(\vec{x},\vec{p},t)=\dot{\vec{\lambda}}\cdot\vec{f}(\vec{x},\vec{p},\vec{\lambda})$
with $\vec{f}(\vec{x},\vec{p},\vec{\lambda})$ depending on
\begin{equation}
\sum_{i}[\frac{\gamma}{\beta}\frac{\partial^{2}H_{\mathrm{a}}}{\partial p_{i}^{2}}-\gamma p_{i}\frac{\partial H_{\mathrm{a}}}{\partial p_{i}}+\frac{\partial H_{\mathrm{a}}}{\partial p_{i}}\frac{\partial H_{\mathrm{o}}}{\partial x_{i}}-p_{i}\frac{\partial H_{\mathrm{a}}}{\partial x_{i}}]=\frac{dF}{dt}-\frac{\partial H_{\mathrm{o}}}{\partial t}.\label{eq:ffunequmove}
\end{equation}
The boundary conditions for the auxiliary Hamiltonian $H_{\mathrm{a}}(t)$
are presented explicitly as $\dot{\vec{\lambda}}(0)=\dot{\vec{\lambda}}(\tau)=0.$
The irreversible energy cost $W_{\mathrm{irr}}\equiv W-\Delta F$
in the shortcut scheme follows as \citep{Li2022}
\begin{equation}
W_{\mathrm{irr}}=\sum_{\mu\nu}\int_{0}^{\tau}dt\dot{\lambda}_{\mu}\dot{\lambda}_{\nu}g_{\mu\nu},\label{eq:irrenergy}
\end{equation}
where the positive semi-definite metric is $g_{\mu\nu}=\gamma\sum_{i}\langle(\partial f_{\mu}/\partial p_{i})(\partial f_{\nu}/\partial p_{i})\rangle_{\mathrm{eq}}$
with $\langle\cdot\rangle_{\mathrm{eq}}=\iint d\vec{x}d\vec{p}[\cdot]P_{\mathrm{eq}}.$
Here $W\equiv\langle\int_{0}^{\tau}dt\partial_{t}H\rangle$ is the
mean work with $\langle\cdot\rangle$ representing the ensemble average
over stochastic trajectories and $\Delta F\equiv F(\vec{\lambda}(\tau))-F(\vec{\lambda}(0))$
is the free energy difference. The metric $g_{\mu\nu}$ endows a Riemannian
manifold in the space of thermodynamic equilibrium states marked by
the control parameters $\vec{\lambda}.$ Minimizing the irreversible
work in Eq.$\ $(\ref{eq:irrenergy}) is equivalent to finding the
geodesic path in the geometric space with the metric $g_{\mu\nu}$.
This property allows us to obtain the optimal control protocol in
the shortcut scheme by using methods developed in geometry \citep{Berger2007}.

Generally, the auxiliary Hamiltonian $H_{\mathrm{a}}$ are momentum-dependent
that are hard to be implemented. For example, the auxiliary Hamiltonian
for a one-dimensional harmonic system $H_{\mathrm{o}}=p^{2}/(2m)+\lambda x^{2}/2$
is obtained as \citep{Li2017,Li2022} $H_{\mathrm{a}}=\dot{\lambda}[(p-\gamma x)^{2}+m\lambda x^{2}]/(4\gamma\lambda).$
The quadratic momentum-dependent term $p^{2}$ and the linear momentum-dependent
term $xp$ in the auxiliary Hamiltonian are hard to be realized in
experiment due to the requirement of constantly monitoring the momentum
\citep{GueryOdelin2022}.

\section{Approximate shortcut scheme\label{SSec-three}}

The variational shortcut scheme is an approximation of the exact shortcut
scheme. The auxiliary Hamiltonian $H_{\mathrm{a}}$ in the exact shortcut
scheme is replaced by the approximate auxiliary Hamiltonian $H_{\mathrm{a}}^{*}$.
We define a semi-positive functional as \citep{Li2021}
\begin{eqnarray}
\mathcal{G}(H_{\mathrm{a}}^{*}) & = & \int d\vec{x}d\vec{p}[\sum_{i}(\frac{\gamma}{\beta}\frac{\partial^{2}H_{\mathrm{a}}^{*}}{\partial p_{i}^{2}}-\gamma p_{i}\frac{\partial H_{\mathrm{a}}^{*}}{\partial p_{i}}+\frac{\partial H_{\mathrm{o}}}{\partial x_{i}}\frac{\partial H_{\mathrm{a}}^{*}}{\partial p_{i}}\nonumber \\
 &  & -p_{i}\frac{\partial H_{\mathrm{a}}^{*}}{\partial x_{i}})+\frac{\partial H_{\mathrm{o}}}{\partial t}-\frac{\mathrm{d}F}{\mathrm{d}t}]^{2}\mathrm{e}^{-\beta H_{\mathrm{o}}}\geqslant0.\label{eq:noncons}
\end{eqnarray}
Finding the exact auxiliary Hamiltonian $H_{\mathrm{a}}^{*}=H_{\mathrm{a}}$
is equivalent to solving the variational equation \citep{Li2021}
\begin{equation}
\frac{\delta\mathcal{G}(H_{\mathrm{a}}^{*})}{\delta H_{\mathrm{a}}^{*}}=0.\label{seq:gaussvar}
\end{equation}
Instead of finding the exact solution, we use the above variational
equation in Eq.$\ (\ref{seq:gaussvar})$ to solve for the possible
approximate Hamiltonian $H_{\mathrm{a}}^{*}$ by finding the minimum
value of $\mathcal{G}(H_{\mathrm{a}}^{*})$ in Eq.$\ $(\ref{eq:noncons}).
With the current variational method, we are able to neglect the quadratic
term $p^{2}$ and remove the linear term $p$. The procedure is divided
into two steps to remove the $p^{2}$ term with approximation and
$p$ terms with gauge transformation, illustrated in Fig.$\ \ref{Schematic}$
with the example Hamiltonian. The details are presented as follows.

\textbf{Step I: Approximate auxiliary control without the quadratic
term $p^{2}$}. The first task is to remove the quadratic term $p^{2}$,
by assuming the form of the auxiliary Hamiltonian
\begin{equation}
H_{\mathrm{a}}^{*}=\sum_{\mu i}\dot{\lambda}_{\mu}B_{\mu}(\vec{\lambda})x_{i}p_{i}+\sum_{\mu i}\dot{\lambda}_{\mu}C_{\mu i}(\vec{\lambda})p_{i}+\sum_{\mu}\dot{\lambda}_{\mu}D_{\mu}(\vec{x},\vec{\lambda}),\label{eq:apphamton}
\end{equation}
where $\vec{B}(\vec{\lambda}),$ $C(\vec{\lambda}),$ and $\vec{D}(\vec{x},\vec{\lambda})$
are functions determined by the variational equation$\ $(\ref{seq:gaussvar}).
Such approximation is valid for the case where the kinetic energy
is negligible in the total energy. We illustrate how such approximation
works with an example in the Appendix$\ $\ref{SAec:A}.

With such approximation, a distribution $P_{\mathrm{eq}}^{*}$ is
reached as an approximation of the instantaneous equilibrium distribution
$P_{\mathrm{eq}}$ with the variational shortcut scheme under the
total Hamiltonian $H^{*}=H_{\mathrm{o}}+H_{\mathrm{a}}^{*}$, i.e.,
$P_{\mathrm{eq}}^{*}\approx P_{\mathrm{eq}}$. The mean work in the
variational shortcut scheme follows as
\begin{eqnarray}
W & = & \langle\int_{0}^{\tau}dt\frac{\partial H_{\mathrm{o}}}{\partial t}\rangle_{\mathrm{eq}}^{*}+\langle\int_{0}^{\tau}dt\frac{\partial H_{\mathrm{a}}^{*}}{\partial t}\rangle_{\mathrm{eq}}^{*}\nonumber \\
 & \approx & \Delta F^{*}+\gamma\sum_{i}\int_{0}^{\tau}dt\iint d\vec{x}d\vec{p}(\frac{\partial H_{\mathrm{a}}^{*}}{\partial p_{i}})^{2}P_{\mathrm{eq}}^{*},\label{eq:approxtwork}
\end{eqnarray}
where $\langle\cdot\rangle_{\mathrm{eq}}^{*}=\iint d\vec{x}d\vec{p}[\cdot]P_{\mathrm{eq}}^{*}.$
The free energy difference $\Delta F^{*}=\langle\int_{0}^{\tau}dt\partial_{t}H_{\mathrm{o}}\rangle_{\mathrm{eq}}^{*}$
is treated as an approximation to the free energy difference $\Delta F$
with high precision \citep{Li2021}. The additional energy cost of
the variational shortcut scheme is evaluated by the irreversible work
as
\begin{eqnarray}
W_{\mathrm{irr}} & \equiv & W-\Delta F\nonumber \\
 & \approx & \gamma\sum_{\mu\nu i}\int_{0}^{\tau}dt\dot{\lambda}_{\mu}\dot{\lambda}_{\nu}\langle\frac{\partial f_{\mu}^{*}}{\partial p_{i}}\frac{\partial f_{\nu}^{*}}{\partial p_{i}}\rangle_{\mathrm{eq}}^{*},\label{eq:appirrwork}
\end{eqnarray}
where $H_{\mathrm{a}}^{*}=\dot{\vec{\lambda}}\cdot\vec{f}^{*}$ with
$\vec{f}^{*}$ representing an approximation to the function $\vec{f}$
in Eq$\ $(\ref{eq:ffunequmove}). With a rescaling of the time $s=t/\tau,$
the irreversible work in Eq.$\ $(\ref{eq:appirrwork}) is proved
to follow the $1/\tau$ scaling which has been widely investigated
in finite-time studies \citep{Curzon1975,Salamon1983,Broeck2005,Schmiedl2007,Esposito2010,Wang2012,Ryabov2016,Ma2018,Ma2018a,Ma2020,Tu2021}.
With the definition of a positive semi-definite metric
\begin{equation}
g_{\mu\nu}^{*}=\gamma\sum_{i}\langle\frac{\partial f_{\mu}^{*}}{\partial p_{i}}\frac{\partial f_{\nu}^{*}}{\partial p_{i}}\rangle_{\mathrm{eq}}^{*},\label{eq:appgeome}
\end{equation}
we can construct a Riemannian manifold on the space of the control
parameters. The shortest curve connecting two equilibrium states is
described through the thermodynamic length \citep{Salamon1983,Crooks2007,Sivak2012,Scandi2019,Chen2021}
$\mathcal{L}=\int_{0}^{\tau}dt\sqrt{\dot{\lambda}_{\mu}\dot{\lambda}_{\nu}g_{\mu\nu}^{*}}$
which gives a lower bound for the irreversible work $W_{\mathrm{irr}}$
as \citep{Crooks2007}
\begin{eqnarray}
W_{\mathrm{irr}} & \ge & \frac{\mathcal{L}^{2}}{\tau}.\label{eq:causchineq}
\end{eqnarray}
Given boundary conditions $\vec{\lambda}(0)$ and $\vec{\lambda}(\tau),$
the lower bound in Eq.$\ $(\ref{eq:causchineq}) is reached by the
optimal control scheme obtained by solving the geodesic equation 
\begin{eqnarray}
\ddot{\lambda}_{\mu}+\sum_{\nu\kappa}\Gamma_{\nu\kappa}^{\mu}\dot{\lambda}_{\nu}\dot{\lambda}_{\kappa} & = & 0,\label{eq:geodesiceq}
\end{eqnarray}
where $\Gamma_{\nu\kappa}^{\mu}\equiv\sum_{\iota}(g^{*-1})_{\iota\mu}(\partial_{\lambda_{\kappa}}g_{\iota\nu}^{*}+\partial_{\lambda_{\nu}}g_{\iota\kappa}^{*}-\partial_{\lambda_{\iota}}g_{\nu\kappa}^{*})/2$
is the Christoffel symbol.

\begin{figure}[!htp]
\includegraphics[width=8.5cm]{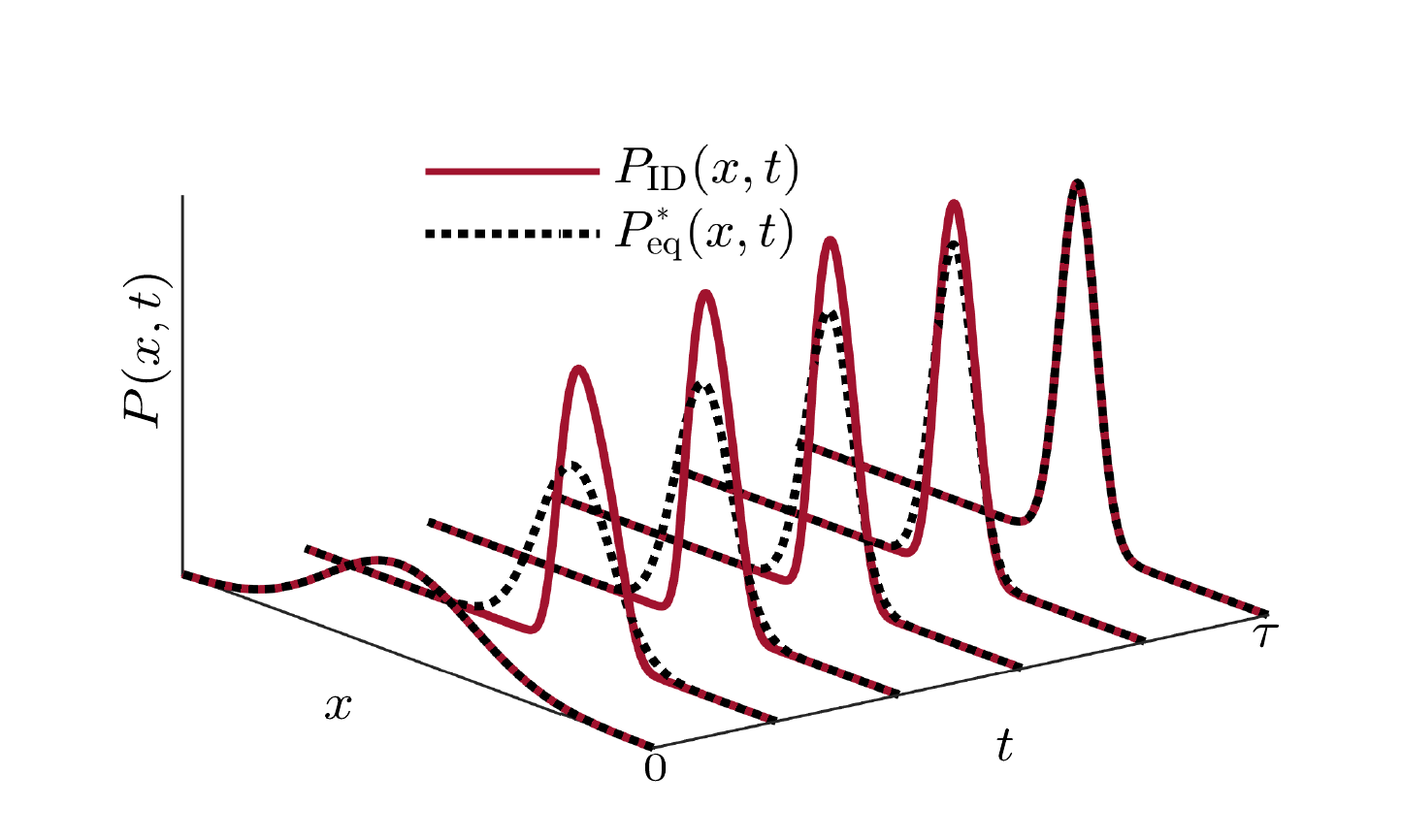} \caption{(Color online) Schematic of processes controlled by the variational
shortcut scheme and the gauge transformation scheme. The system distribution
of the variational shortcut scheme keeps in instantaneous equilibrium
$P_{\mathrm{eq}}^{*}(x,t)$ during the evolution process while the
instantaneous distribution of the gauge transformation scheme $P_{\mathrm{ID}}(x,t)$
deviates from the equilibrium distribution $P_{\mathrm{eq}}^{*}(x,t)$
for intermediate times $(0<t<\tau)$.}
\label{Distribution}
\end{figure}

\textbf{Step II: Equivalent process without the linear term $p$}.
The higher-order terms of the momentum in Eq.$\ (\ref{eq:apphamton})$
are removed for the experimental feasibility. In the variational shortcut
scheme, the dynamical evolution of the system with the Hamiltonian
$H=H_{\mathrm{o}}+H_{\mathrm{a}}^{*}$ is governed by the Langevin
equation, 
\begin{multline}
\dot{x}_{i}=\frac{p_{i}}{m}+\sum_{\mu}\dot{\lambda}_{\mu}B_{\mu}x_{i}+\sum_{\mu}\dot{\lambda}_{\mu}C_{\mu i},\\
\dot{p}_{i}=-\frac{\partial U_{\mathrm{o}}}{\partial x_{i}}-\sum_{\mu}\dot{\lambda}_{\mu}B_{\mu}p_{i}-\sum_{\mu}\dot{\lambda}_{\mu}\frac{\partial D_{\mu}}{\partial x_{i}}-\gamma\dot{x}_{i}+\xi_{i}(t),\label{eq:degenlaneq}
\end{multline}
where $\vec{\xi}\equiv(\xi_{1},\xi_{2},\cdots,\xi_{N})$ are the Gaussian
white noise. During the evolution process described by Eq.$\ $(\ref{eq:degenlaneq}),
the distribution of the system always stays in the instantaneous equilibrium
distribution 
\begin{equation}
P_{\mathrm{eq}}^{*}(\vec{x},\vec{p},\vec{\lambda})=\exp\{\beta[F(\vec{\lambda})-\sum_{i}p_{i}^{2}/(2m)-U_{\mathrm{o}}(\vec{x},\vec{\lambda})]\}.\label{eq:insequdist}
\end{equation}
With a gauge transformation \citep{Sels2017,Li2021} $p_{i}\to p_{i}+m\sum_{\mu}\dot{\lambda}_{\mu}B_{\mu}x_{i}+m\sum_{\mu}\dot{\lambda}_{\mu}C_{\mu i},$
we can obtain an equivalent process controlled by the original Hamiltonian
$H_{\mathrm{o}}$ and the auxiliary force $\vec{F}^{\mathrm{a}}$

\begin{eqnarray}
\dot{x}_{i} & = & \frac{p_{i}}{m},\nonumber \\
\dot{p}_{i} & = & -\frac{\partial U_{\mathrm{o}}}{\partial x_{i}}+F_{i}^{\mathrm{a}}-\gamma\dot{x}_{i}+\xi_{i}(t),\label{eq:afterchangevar}
\end{eqnarray}
where the auxiliary force is explicitly presented as 
\begin{eqnarray}
F_{i}^{\mathrm{a}} & = & -\sum_{\mu}\dot{\lambda}_{\mu}\frac{\partial D_{\mu}}{\partial x_{i}}+m\sum_{\mu}\ddot{\lambda}_{\mu}B_{\mu}x_{i}+m\sum_{\mu\nu}\dot{\lambda}_{\mu}\dot{\lambda}_{\nu}\frac{\partial B_{\mu}}{\partial\lambda_{\nu}}x_{i}\nonumber \\
 &  & +m\sum_{\mu\nu}\dot{\lambda}_{\mu}\dot{\lambda}_{\nu}B_{\mu}B_{\nu}x_{i}+m\sum_{\mu\nu}\dot{\lambda}_{\mu}\dot{\lambda}_{\nu}B_{\mu}C_{\nu i}\nonumber \\
 &  & +m\sum_{\mu}\ddot{\lambda}_{\mu}C_{\mu i}+m\sum_{\mu\nu}\dot{\lambda}_{\mu}\dot{\lambda}_{\nu}\frac{\partial C_{\mu i}}{\partial\lambda_{\nu}}.\label{eq:equivforce}
\end{eqnarray}
The distribution of the system in the process described by Eq.$\ $(\ref{eq:afterchangevar})
keeps in an instantaneous distribution with the fixed pattern
\begin{eqnarray}
P_{\mathrm{ID}}(\vec{x},\vec{p},\vec{\lambda}) & = & \exp\{\beta[F(\vec{\lambda})-\sum_{i}(p_{i}-m\sum_{\mu}\dot{\lambda}_{\mu}B_{\mu}x_{i}\nonumber \\
 &  & -m\sum_{\mu}\dot{\lambda}_{\mu}C_{\mu i})^{2}/(2m)-U_{\mathrm{o}}(\vec{x},\vec{\lambda})]\}.\label{eq:insdisf}
\end{eqnarray}
As shown in Fig$\ $\ref{Distribution}, with the boundary condition
$\dot{\vec{\lambda}}(0)=\dot{\vec{\lambda}}(\tau)=0$, the instantaneous
distribution $P_{\mathrm{ID}}(\vec{x},\vec{p},\vec{\lambda})$ in
Eq.$\ $(\ref{eq:insdisf}) returns to the instantaneous equilibrium
distribution $P_{\mathrm{eq}}^{*}(\vec{x},\vec{p},\vec{\lambda})$
in Eq.$\ $(\ref{eq:insequdist}) at $t=0$ and $\tau.$ It means
that the system following the Langevin equation$\ $(\ref{eq:afterchangevar})
can evolve from an initial equilibrium state to another target equilibrium
state. The operation of gauge transformation does not change the irreversible
work in Eq.$\ $(\ref{eq:appirrwork}). Therefore, the optimal control
from the geodesic equation$\ $(\ref{eq:geodesiceq}) applies equally
to the process described by the Langevin equation$\ $(\ref{eq:afterchangevar}).
For the optimized protocol in the shortcut scheme, the boundary condition
for $\dot{\vec{\lambda}}(t)$ is usually realized as discrete jumps
at the beginning and end of the auxiliary force \citep{Li2022}. In
the gauge transformation scheme, the system distribution in Eq.$\ $(\ref{eq:insdisf})
explicitly depends on $\dot{\vec{\lambda}}(t)$. And there is a mismatching
between the initial equilibrium state $P_{\mathrm{eq}}^{*}(\vec{x},\vec{p},\vec{\lambda}(0))$
and the system distribution $P_{\mathrm{ID}}(\vec{x},\vec{p},\vec{\lambda}(0)),$
which can be offset for systems with weak inertial effect. We illustrate
this claim in the following example.

The steps to derive the experimentally testable protocol with minimum
energy cost are shown in Fig.$\ \ref{Schematic}$. In step I, we solve
for the best possible auxiliary Hamiltonian $H_{\mathrm{a}}^{*}$
in Eq.$\ $(\ref{seq:gaussvar}) by using the variational shortcut
scheme. The geometric approach is then applied to minimize the irreversible
work $W_{\mathrm{irr}}$ in Eq.$\ $(\ref{eq:appirrwork}) and obtain
the optimal protocol. In step II, the gauge transformation scheme
with the operation in Eq.$\ $(\ref{eq:afterchangevar}) is carried
out to obtain the final experimentally testable protocol with minimum
energy cost.

\section{Application\label{SSec-four}}

To demonstrate our strategy, we consider the Brownian motion in the
harmonic potential with two controllable parameters $\vec{\lambda}=(\lambda_{1},\lambda_{2})$
with Hamiltonian as 
\begin{eqnarray}
H_{\mathrm{o}} & = & \frac{p^{2}}{2m}+\frac{\lambda_{1}}{2}x^{2}-\lambda_{2}x.\label{eq:originalH}
\end{eqnarray}
In the shortcut scheme \citep{Li2017,Li2022}, the exact auxiliary
Hamiltonian takes the form $H_{\mathrm{a}}(x,p,t)=\sum_{\mu=1}^{2}\dot{\lambda}_{\mu}f_{\mu}(x,p,\lambda_{1},\lambda_{2})$
where
\begin{eqnarray}
f_{1} & = & \frac{(p-\gamma x)^{2}+m\lambda_{1}x^{2}}{4\gamma\lambda_{1}}-\frac{\lambda_{2}p}{2\lambda_{1}^{2}}+(\frac{\gamma\lambda_{2}}{2\lambda_{1}^{2}}-\frac{m\lambda_{2}}{2\gamma\lambda_{1}})x,\nonumber \\
f_{2} & = & \frac{p}{\lambda_{1}}-\frac{\gamma x}{\lambda_{1}}.\label{eq:auxham}
\end{eqnarray}
With the application of the geometric approach proposed in Ref. \citep{Li2022},
the geodesic (optimal) protocol with minimum energy cost can be obtained
as 
\begin{eqnarray}
\dot{\lambda}_{1} & = & \frac{w}{\tau}\sqrt{\frac{\lambda_{1}^{3}}{\lambda_{1}+\gamma^{2}/m}},\nonumber \\
\frac{\lambda_{2}}{\lambda_{1}} & = & a\frac{t}{\tau}+b,\label{eq:geoprotocol}
\end{eqnarray}
where $w=-\{2\sqrt{1+\gamma^{2}/(m\lambda_{1})}+\ln[\sqrt{1+\gamma^{2}/(m\lambda_{1})}-1]-\ln[\sqrt{1+\gamma^{2}/(m\lambda_{1})}+1]\}\mid_{\lambda_{1}(0)}^{\lambda_{1}(\tau)}$,
$a=[\lambda_{2}(\tau)\lambda_{1}(0)-\lambda_{2}(0)\lambda_{1}(\tau)]/[\lambda_{1}(\tau)\lambda_{1}(0)]$,
and $b=\lambda_{2}(0)/\lambda_{1}(0)$ are constants. Here $\vec{\lambda}(0)=(\lambda_{1}(0),\lambda_{2}(0))$
and $\vec{\lambda}(\tau)=(\lambda_{1}(\tau),\lambda_{2}(\tau))$ are
boundary conditions. The momentum-dependent terms in Eq.$\ $(\ref{eq:auxham})
hinder the implementation of the shortcut scheme in experiment.

We assume that the approximate auxiliary Hamiltonian takes the form
\begin{equation}
H_{\mathrm{a}}^{*}=a_{1}(t)px+a_{2}(t)p+a_{3}(t)x^{2}+a_{4}(t)x,\label{seq:appham}
\end{equation}
 where $a_{1}(t),$ $a_{2}(t),$ $a_{3}(t),$ and $a_{4}(t)$ are
coefficients to be determined. And the variational functional in Eq.$\ $(\ref{eq:noncons})
follows as 
\begin{eqnarray}
\mathcal{G} & = & \iint dxdp\{[a_{1}p^{2}+(\gamma a_{1}+2a_{3})xp+(\gamma a_{2}+a_{4})p\nonumber \\
 &  & -m(a_{1}x+a_{2})(\lambda_{1}x-\lambda_{2})]^{2}+\frac{2}{\beta}(a_{1}x+a_{2})(\dot{\lambda}_{1}x-\dot{\lambda}_{2})\}\nonumber \\
 &  & \times\exp[-\beta(\frac{p^{2}}{2m}+\frac{\lambda_{1}}{2}x^{2}-\lambda_{2}x)].\label{eq:noneqcons}
\end{eqnarray}
The best possible auxiliary Hamiltonian $H_{\mathrm{a}}^{*}$ is obtained
by minimizing the variational functional $\mathcal{G}$ over the parameters
$a_{1}(t),$ $a_{2}(t),$ $a_{3}(t),$ and $a_{4}(t)$ as
\begin{equation}
H_{\mathrm{a}}^{*}=\dot{\lambda}_{1}(-\frac{px}{4\lambda_{1}}-\frac{3\lambda_{2}p}{4\lambda_{1}^{2}}+\frac{\gamma x^{2}}{8\lambda_{1}}+\frac{3\gamma\lambda_{2}x}{4\lambda_{1}^{2}})+\dot{\lambda}_{2}(\frac{p}{\lambda_{1}}-\frac{\gamma x}{\lambda_{1}})\text{.}\label{seq:approxham}
\end{equation}
Detailed calculations are presented in Appendix$\ $\ref{SAec:A}.

We then apply the geometric approach to derive the optimal protocol
with minimum energy cost. With the auxiliary Hamiltonian in Eq.$\ $(\ref{seq:approxham}),
the geometric metric in Eq.$\ $(\ref{eq:appgeome}) follows as 
\begin{eqnarray}
g & = & \left(\begin{array}{cc}
\frac{\gamma}{16\beta\lambda_{1}^{3}}+\frac{\gamma\lambda_{2}^{2}}{\lambda_{1}^{4}} & -\frac{\gamma\lambda_{2}}{\lambda_{1}^{3}}\\
-\frac{\gamma\lambda_{2}}{\lambda_{1}^{3}} & \frac{\gamma}{\lambda_{1}^{2}}
\end{array}\right).\label{eq:metric}
\end{eqnarray}
The geodesic equation with the metric in Eq.$\ $(\ref{eq:metric})
takes the form
\begin{eqnarray}
 &  & \ddot{\lambda}_{1}-\frac{3\dot{\lambda}_{1}^{2}}{2\lambda_{1}}=0,\nonumber \\
 &  & \ddot{\lambda}_{2}-\frac{2\dot{\lambda}_{1}\dot{\lambda}_{2}}{\lambda_{1}}+\frac{\dot{\lambda}_{1}^{2}\lambda_{2}}{2\lambda_{1}^{2}}=0.\label{eq:geodesicequa}
\end{eqnarray}
The solution for Eq.$\ $(\ref{eq:geodesicequa}) is analytically
obtained as
\begin{eqnarray}
\lambda_{1}(t) & = & \frac{1}{[(\lambda_{1}^{-\frac{1}{2}}(\tau)-\lambda_{1}^{-\frac{1}{2}}(0))\frac{t}{\tau}+\lambda_{1}^{-\frac{1}{2}}(0)]^{2}},\nonumber \\
\lambda_{2}(t) & = & \frac{(\frac{\lambda_{2}(\tau)}{\lambda_{1}(\tau)}-\frac{\lambda_{2}(0)}{\lambda_{1}(0)})\frac{t}{\tau}+\frac{\lambda_{2}(0)}{\lambda_{1}(0)}}{[(\lambda_{1}^{-\frac{1}{2}}(\tau)-\lambda_{1}^{-\frac{1}{2}}(0))\frac{t}{\tau}+\lambda_{1}^{-\frac{1}{2}}(0)]^{2}}.\label{eq:anageodeseq}
\end{eqnarray}

\begin{figure}[!htp]
\includegraphics[width=8.5cm]{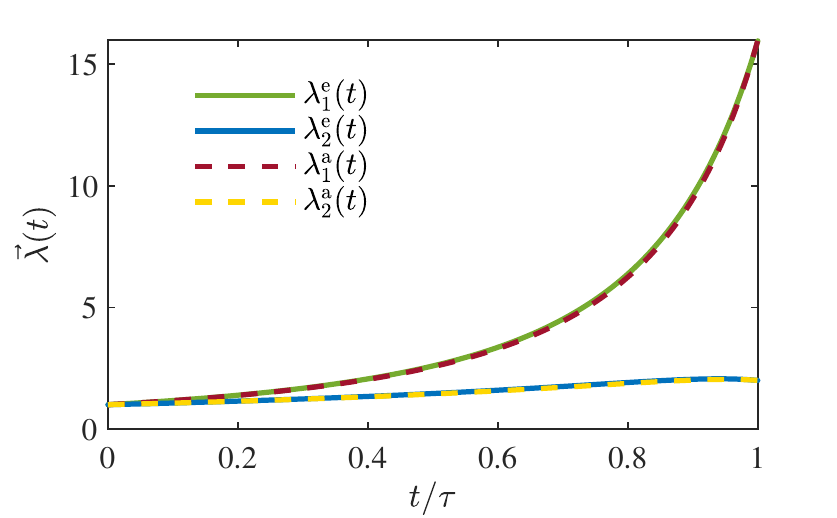} \caption{(Color online) Geodesic protocols for the exact shortcut scheme and
the variational shortcut scheme. We have set the parameters as $m=0.01,$
$\gamma=1,$ and $k_{\mathrm{B}}T=1.$ The protocols change from $\vec{\lambda}(0)=(1,1)$
to $\vec{\lambda}(\tau)=(16,2).$ The solid lines are the geodesic
protocol $\vec{\lambda}^{\mathrm{e}}(t)$ for the exact shortcut scheme
in Eq.$\ (\ref{eq:geoprotocol})$ and the dashed lines are the geodesic
protocol $\vec{\lambda}^{\mathrm{a}}(t)$ for the variational shortcut
scheme in Eq.$\ (\ref{eq:anageodeseq})$. The geodesic protocol for
the variational shortcut scheme (dashed lines) keeps close to the
one for the exact shortcut scheme (solid lines).}
\label{Protocol}
\end{figure}
In Fig.$\ $\ref{Protocol}, we compare geodesic protocols for the
exact shortcut scheme and the variational shortcut scheme. The parameters
are chosen as $\vec{\lambda}(0)=(1,1),$ $\vec{\lambda}(\tau)=(16,2),$
$m=0.01,$ $\gamma=1,$ and $k_{\mathrm{B}}T=1.$ The geodesic protocol
for the variational shortcut scheme $\vec{\lambda}^{\mathrm{a}}(t)$
(dashed lines) is in close proximity to the one for the exact shortcut
scheme $\vec{\lambda}^{\mathrm{e}}(t)$ (solid lines), which supports
our claim that the variational shortcut scheme can approximately reproduce
the results of the exact shortcut scheme.

Note that there are still linear terms of the momentum in Eq.$\ $(\ref{seq:approxham}).
In the second step, we remove the remaining momentum-dependent terms
by using the gauge transformation scheme. The Langevin equation for
the variational shortcut scheme with the Hamiltonian $H=H_{\mathrm{o}}+H_{\mathrm{a}}^{*}$
follows as
\begin{multline}
\dot{x}=\frac{p}{m}-\frac{\dot{\lambda}_{1}x}{4\lambda_{1}}-\frac{3\dot{\lambda}_{1}\lambda_{2}}{4\lambda_{1}^{2}}+\frac{\dot{\lambda}_{2}}{\lambda_{1}},\\
\dot{p}=-\lambda_{1}x+\lambda_{2}+\frac{\dot{\lambda}_{1}p}{4\lambda_{1}}-\frac{\gamma\dot{\lambda}_{1}x}{4\lambda_{1}}-\frac{3\gamma\dot{\lambda}_{1}\lambda_{2}}{4\lambda_{1}^{2}}+\frac{\gamma\dot{\lambda}_{2}}{\lambda_{1}}-\gamma\dot{x}+\xi(t).\label{seq:applange}
\end{multline}
With the gauge transformation
\begin{equation}
p\to p-\frac{m\dot{\lambda}_{1}x}{4\lambda_{1}}-\frac{3m\dot{\lambda}_{1}\lambda_{2}}{4\lambda_{1}^{2}}+\frac{m\dot{\lambda}_{2}}{\lambda_{1}},\label{seq:changevariable}
\end{equation}
 the Langevin equation$\ $(\ref{seq:applange}) is transformed into
\begin{eqnarray}
\dot{x} & = & p,\nonumber \\
\dot{p} & = & -\lambda_{1}x+\lambda_{2}+F_{\mathrm{a}}-\gamma p+\xi(t),\label{seq:alternadyna}
\end{eqnarray}
where the auxiliary force follows as
\begin{eqnarray}
F_{\mathrm{a}} & \equiv & (\frac{5m\dot{\lambda}_{1}^{2}}{16\lambda_{1}^{2}}-\frac{\gamma\dot{\lambda}_{1}}{4\lambda_{1}}-\frac{m\ddot{\lambda}_{1}}{4\lambda_{1}})x-\frac{3\gamma\dot{\lambda}_{1}\lambda_{2}}{4\lambda_{1}^{2}}+\frac{\gamma\dot{\lambda}_{2}}{\lambda_{1}}\nonumber \\
 &  & +\frac{27m\dot{\lambda}_{1}^{2}\lambda_{2}}{16\lambda_{1}^{3}}-\frac{2m\dot{\lambda}_{1}\dot{\lambda}_{2}}{\lambda_{1}^{2}}-\frac{3m\ddot{\lambda}_{1}\lambda_{2}}{4\lambda_{1}^{2}}+\frac{m\ddot{\lambda}_{2}}{\lambda_{1}}.\label{eq:auxiliforce}
\end{eqnarray}
In the dynamics governed by the transformed Langevin equation$\ $(\ref{seq:alternadyna}),
there is no momentum-dependent term in the driving force. The auxiliary
potential corresponding to the driving force in Eq.$\ $(\ref{eq:auxiliforce})
is obtained as
\begin{eqnarray}
U_{\mathrm{a}} & = & \frac{m\ddot{\lambda}_{1}}{8\lambda_{1}}x^{2}+\frac{\gamma\dot{\lambda}_{1}}{8\lambda_{1}}x^{2}-\frac{5m\dot{\lambda}_{1}^{2}}{32\lambda_{1}^{2}}x^{2}-\frac{m\ddot{\lambda}_{2}}{\lambda_{1}}x-\frac{27m\dot{\lambda}_{1}^{2}\lambda_{2}}{16\lambda_{1}^{3}}x\nonumber \\
 &  & +\frac{2m\dot{\lambda}_{1}\dot{\lambda}_{2}}{\lambda_{1}^{2}}x+\frac{3m\ddot{\lambda}_{1}\lambda_{2}}{4\lambda_{1}^{2}}x-\frac{\gamma\dot{\lambda}_{2}}{\lambda_{1}}x+\frac{3\gamma\dot{\lambda}_{1}\lambda_{2}}{4\lambda_{1}^{2}}x.\label{eq:auxpotential}
\end{eqnarray}
 The system can be approximately transformed from an initial equilibrium
state to another one within finite time. During intermediate driving
process, the system follows the instantaneous distribution
\begin{eqnarray}
P_{\mathrm{ID}}(x,p,\vec{\lambda}) & = & \exp\{\beta[F(\vec{\lambda})-\frac{1}{2m}(p+\frac{m\dot{\lambda}_{1}}{4\lambda_{1}}x\nonumber \\
 &  & +\frac{3m\dot{\lambda}_{1}\lambda_{2}}{4\lambda_{1}^{2}}-\frac{m\dot{\lambda}_{2}}{\lambda_{1}})^{2}-U_{\mathrm{o}}(x,\vec{\lambda})]\},\label{eq:insdisf-1}
\end{eqnarray}
which is an approximation to the instantaneous equilibrium states
$P_{\mathrm{eq}}$.

To validate the variational shortcut scheme and the gauge transformation
scheme, we compare the distributions $P_{\mathrm{eq}}^{*}(x,p,\vec{\lambda})$
and $P_{\mathrm{ID}}(x,p,\vec{\lambda})$ with that of the exact shortcut
scheme $P_{\mathrm{eq}}(x,p,\vec{\lambda})$. The variational shortcut
scheme and the gauge transformation scheme are respectively implemented
through the Hamiltonian $H_{\mathrm{o}}+H_{\mathrm{a}}^{*}$ and $H_{\mathrm{o}}+U_{\mathrm{a}}$
while the exact shortcut scheme is realized through the Hamiltonian
$H_{\mathrm{o}}+H_{\mathrm{a}}$. The distance between the instantaneous
equilibrium distribution $P_{\mathrm{eq}}$ and the distribution in
the variational shortcut scheme $P=P_{\mathrm{eq}}^{*}$ or the distribution
in the gauge transformation scheme $P=P_{\mathrm{ID}}$ are evaluated
through the Jensen-Shannon divergence \citep{Lin1991,Endres2003,Majtey2005,Feng2008}
\begin{equation}
\mathrm{D}(P||P_{\mathrm{eq}})=\frac{1}{2}\int d\vec{x}d\vec{p}(P\ln\frac{2P}{P+P_{\mathrm{eq}}}+P_{\mathrm{eq}}\ln\frac{2P_{\mathrm{eq}}}{P+P_{\mathrm{eq}}}).\label{eq:JSdistance}
\end{equation}

\begin{figure}[!htp]
\includegraphics[width=8.5cm]{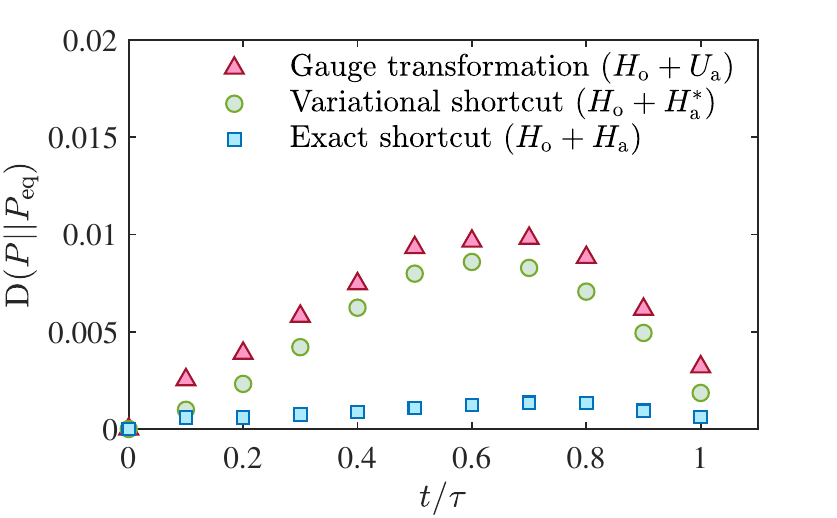} \caption{(Color online) Distance $\mathrm{D}(P||P_{\mathrm{eq}})$ between
the instantaneous equilibrium distribution $P_{\mathrm{eq}}$ and
the distribution in the variational shortcut scheme $P=P_{\mathrm{eq}}^{*}$
or the distribution in the gauge transformation scheme $P=P_{\mathrm{ID}}$.
The gauge transformation scheme (red triangles), the variational shortcut
scheme (green circles), and the exact shortcut scheme (blue squares)
are implemented through the Hamiltonian $H_{\mathrm{o}}+U_{\mathrm{a}},$
$H_{\mathrm{o}}+H_{\mathrm{a}}^{*},$ and $H_{\mathrm{o}}+H_{\mathrm{a}},$
respectively. The distance is evaluated by using the Jensen-Shannon
divergence \citep{Lin1991,Endres2003,Majtey2005,Feng2008}. The protocol
$\vec{\lambda}^{\mathrm{e}}(t)$ in Eq.$\ $(\ref{eq:geoprotocol})
is chosen to realize different schemes for fair comparison.}
\label{Distance}
\end{figure}

We plot the Jensen-Shannon divergence $\mathrm{D}(P||P_{\mathrm{eq}})$
as a function of evolution time $t$ in Fig.$\ $\ref{Distance}.
The protocol $\vec{\lambda}^{\mathrm{e}}(t)$ in Eq.$\ $(\ref{eq:geoprotocol})
is used to realize different driving schemes. Red triangles, green
circles, and blue squares respectively represent the distance from
the equilibrium distribution to the distribution of the gauge transformation
scheme $(H_{\mathrm{o}}+U_{\mathrm{a}})$, the variational shortcut
scheme $(H_{\mathrm{o}}+H_{\mathrm{a}}^{*})$, and the exact shortcut
scheme $(H_{\mathrm{o}}+H_{\mathrm{a}})$. The distribution of the
exact shortcut scheme closely follows the instantaneous equilibrium
distribution while the distribution of the variational shortcut scheme
initially drives the system away from equilibrium and then gradually
back to the final equilibrium state. Compared with the variational
shortcut scheme, the distribution of the gauge transformation scheme
further departs from the instantaneous equilibrium distribution but
still returns to the target equilibrium state approximately. These
results therefore demonstrate that the gauge transformation scheme
can reconcile the experimental feasibility and the target of transforming
the system to the final equilibrium state with high precision.

\begin{figure}[!htp]
\includegraphics[width=8.5cm]{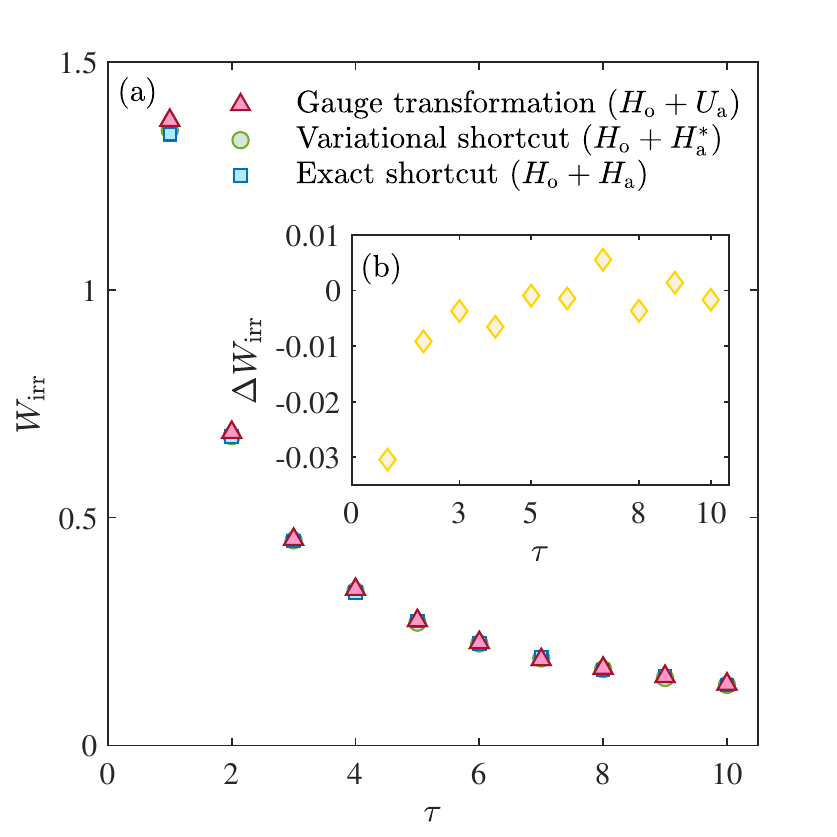} \caption{(Color online) The irreversible work for the gauge transformation
scheme (red triangles), the variational shortcut scheme (green circles),
and the exact shortcut scheme (blue squares) with different durations
$\tau$. The mean work are obtained from $10^{5}$ stochastic trajectories.
The irreversible work of the gauge transformation scheme and the variational
shortcut scheme coincide well, which demonstrates that the gauge transformation
scheme works. The protocol $\vec{\lambda}^{\mathrm{e}}(t)$ in Eq.$\ $(\ref{eq:geoprotocol})
is used to perform the exact shortcut scheme while the protocol $\vec{\lambda}^{\mathrm{a}}(t)$
in Eq.$\ $(\ref{eq:anageodeseq}) is used to realize the variational
shortcut scheme and the gauge transformation scheme. The inset shows
the difference of the irreversible work $\Delta W_{\mathrm{irr}}$
between the exact shortcut scheme and the gauge transformation scheme
for different durations $\tau$.}
\label{Irrwork}
\end{figure}

We also compare the irreversible work $W_{\mathrm{irr}}=W-\Delta F$
for different driving schemes. In the gauge transformation scheme,
the variational shortcut scheme, and the exact shortcut scheme, the
total Hamiltonian $H$ takes the form as $H_{\mathrm{o}}+U_{\mathrm{a}},$
$H_{\mathrm{o}}+H_{\mathrm{a}}^{*},$ and $H_{\mathrm{o}}+H_{\mathrm{a}},$
respectively. The exact shortcut scheme is carried out through the
geodesic protocol $\vec{\lambda}^{\mathrm{e}}(t)$ in Eq.$\ $(\ref{eq:geoprotocol})
while the variational shortcut scheme and the gauge transformation
scheme are carried out through the geodesic protocol $\vec{\lambda}^{\mathrm{a}}(t)$
in Eq.$\ $(\ref{eq:anageodeseq}). Fig.$\ $\ref{Irrwork} shows
the irreversible work of the gauge transformation scheme (red triangles),
the variational shortcut scheme (green circles), and the exact shortcut
scheme (blue squares) for different durations $\tau.$ The mean work
is an average over $10^{5}$ stochastic trajectories. The irreversible
work of the gauge transformation scheme coincides well with that of
the variational shortcut scheme for different driving durations $\tau$,
which demonstrates that the gauge transformation scheme can also well
reproduce the energetic cost of the variational shortcut scheme. The
irreversible work of the exact shortcut scheme is smaller than that
of the gauge transformation scheme in short driving processes and
gradually coincides with later as the duration increases. This shows
that the energy cost of the gauge transformation scheme can keep pace
with that of the exact shortcut scheme while approximately transform
the system to the target equilibrium state with high precision.

\section{Conclusion and discussion\label{SSec-five}}

In conclusion, we have presented a momentum-independent driving scheme
to approximately transform the system from an initial equilibrium
state to another target equilibrium state. The momentum-dependent
terms in the auxiliary Hamiltonian have been removed through the variational
method and the gauge transformation. A geometric approach has been
applied to minimize the energy cost of the driving scheme. The optimal
protocol with minimum energy cost is obtained by solving the geodesic
equation with methods developed in Riemannian geometry. We have tested
our driving strategy by using a Brownian particle system with two
controllable parameters. The simulation results prove that our scheme
can achieve the task of rapidly driving the system to the target equilibrium
state with high precision while reconcile the experimental feasibility.
Our scheme should offer an experimentally testable control protocol
with minimum energy cost for approximately realizing the shortcut
scheme.

The gauge transformation scheme and the variational shortcut scheme
are approximate shortcut schemes. The remaining distance between the
final distribution of the approximate shortcut scheme and the target
equilibrium distribution can be replenished through a short relaxation.
Such a deviation is caused by the absence of the high-order momentum-dependent
terms in the variational shortcut scheme. The momentum-dependent terms
will play an important role in the dynamical evolution if the inertial
effect is significantly obvious. Therefore, the deviation could be
reduced if the inertial effect is weaken. The operation of changing
variables to remove momentum-dependent terms in the gauge transformation
scheme is similar to the fast-forward scheme in the field of shortcuts
to adiabaticity \citep{Masuda2009,Torrontegui2012,MartinezGaraot2016,Jarzynski2017,Patra2017,Sels2017,Kolodrubetz2017,GueryOdelin2019,GueryOdelin2022}.
Our operation is implemented for systems following the stochastic
dynamics.

In the underdamped case, the controlled Brownian motion has been investigated
in different experimental platforms \citep{Li2010,Cunuder2016,Ciliberto2017,Hoang2018,Dago2021}.
The driving force of our approximate shortcut scheme in Eq.$\ $(\ref{eq:auxpotential})
only depends on the position of the system. It is promising to test
our approximate shortcut scheme in experiment.

\section{Acknowledgement\label{SSec-six}}

\emph{Acknowledgement.}--This work is supported by the National Natural
Science Foundation of China (NSFC) (Grants No. 12088101, No. 11875049,
No. U1930402, and No. U1930403) and the National Basic Research Program
of China (Grant No. 2016YFA0301201).

\bibliographystyle{apsrev4-1}
\bibliography{ref}

\begin{widetext}

\appendix

\section{Illustration of how the approximation of the variational shortcut
scheme works\label{SAec:A}}

We illustrate how the approximation of the variational shortcut scheme
works with an example of a Brownian particle moving in the harmonic
potential with the Hamiltonian $H_{\mathrm{o}}=p^{2}/(2m)+\lambda(t)x^{2}/2,$
where $\lambda(t)$ is the control parameter. In the shortcut scheme,
the auxiliary Hamiltonian follows as \citep{Li2017} 
\begin{equation}
H_{\mathrm{a}}=\frac{\dot{\lambda}}{4\gamma\lambda}[(p-\gamma x)^{2}+m\lambda x^{2}].\label{seq:auxconsin}
\end{equation}
To evaluate the contribution of each momentum-dependent term to the
shortcut scheme, we introduce the characteristic length $l_{\mathrm{c}}\equiv(k_{\mathrm{B}}T/\lambda(0))^{1/2}$,
the characteristic time $\tau_{1}=m/\gamma$ and $\tau_{2}=\gamma/\lambda(0)$
to rescale the Hamiltonian. The dimensionless coordinate, momentum,
time, and control protocol are defined as $\tilde{x}\equiv x/l_{\mathrm{c}},$
$\tilde{p}\equiv p\tau_{2}/(ml_{\mathrm{c}})$, $s\equiv t/\tau_{2},$
and $\tilde{\lambda}\equiv\lambda/\lambda(0)$$.$ The dimensionless
Hamiltonian are then obtained as 
\begin{equation}
\tilde{H}_{\mathrm{o}}\equiv\frac{H_{\mathrm{o}}}{k_{\mathrm{B}}T}=\alpha\frac{\tilde{p}^{2}}{2}+\frac{1}{2}\tilde{\lambda}\tilde{x}^{2},\label{seq:origsin}
\end{equation}
and 
\begin{eqnarray}
\tilde{H}_{\mathrm{a}} & \equiv & \frac{H_{\mathrm{a}}}{k_{\mathrm{B}}T}=\frac{\tilde{\lambda}'}{4\tilde{\lambda}}[\alpha^{2}(\tilde{p}-\frac{\tilde{x}}{\alpha})^{2}+\alpha\tilde{\lambda}\tilde{x}^{2}]\nonumber \\
 & = & \alpha^{2}\frac{\tilde{\lambda}'\tilde{p}^{2}}{4\tilde{\lambda}}-\alpha\frac{\tilde{\lambda}'\tilde{x}\tilde{p}}{2\tilde{\lambda}}+\alpha\frac{\tilde{\lambda}'\tilde{x}^{2}}{4}+\frac{\tilde{\lambda}'\tilde{x}^{2}}{4\tilde{\lambda}},\label{seq:auxressin}
\end{eqnarray}
where $\tilde{\lambda}'\equiv d\tilde{\lambda}/ds$ and $\alpha\equiv m\lambda(0)/\gamma^{2}.$
Note that there are different orders of $\alpha$ in the above expressions
of $\tilde{H}_{\mathrm{o}}$ and $\tilde{H}_{\mathrm{a}}.$ If we
assume that $\alpha\ll1$, the second-order term of $\alpha$ in Eq.$\ $(\ref{seq:auxressin}),
i.e., the $\tilde{p}^{2}$ term in $\tilde{H}_{\mathrm{a}}$ can be
neglected, which means that the approximation of the variational shortcut
scheme is valid. The dimensionless parameter $\alpha$ is small if
the mass $m$ or the stiffness coefficient $\lambda(0)$ is small
compared to the dissipation coefficient $\gamma.$

\section{The approximate auxiliary Hamiltonian\label{SAec:B}}

We start from the functional in Eq.$\ $(\ref{eq:noneqcons}). With
the minimization of parameters $a_{1}(t),$ $a_{2}(t),$ $a_{3}(t),$
and $a_{4}(t)$, we obtain a set of equations:

\begin{eqnarray}
 &  & 3a_{1}+\gamma\beta[(\gamma a_{1}+2a_{3})\bar{x^{2}}+(\gamma a_{2}+a_{4})\bar{x}]+\lambda_{1}\beta(a_{1}\bar{x^{2}}+a_{2}\bar{x})+\dot{\lambda}_{1}\beta\bar{x^{2}}-\dot{\lambda}_{2}\beta\bar{x}=0,\nonumber \\
 &  & \gamma a_{2}+a_{4}+(\gamma a_{1}+2a_{3})\bar{x}+\lambda_{1}(a_{1}\bar{x}+a_{2})+\dot{\lambda}_{1}\bar{x}-\dot{\lambda}_{2}=0,\nonumber \\
 &  & (\gamma a_{1}+2a_{3})\bar{x^{2}}+(\gamma a_{2}+a_{4})\bar{x}=0,\nonumber \\
 &  & (\gamma a_{1}+2a_{3})\bar{x}+\gamma a_{2}+a_{4}=0,\label{seq:setequ}
\end{eqnarray}
where $\bar{x^{n}}\equiv\int_{-\infty}^{\infty}x^{n}\exp[-\beta(\lambda_{1}x^{2}/2-\lambda_{2}x)]dx/\int_{-\infty}^{\infty}\exp[-\beta(\lambda_{1}x^{2}/2-\lambda_{2}x)]dx$
with $n=1,2,\cdots.$ We can analytically obtain that $\bar{x}=\lambda_{2}/\lambda_{1},$
$\bar{x^{2}}=1/(\beta\lambda_{1})+\lambda_{2}^{2}/\lambda_{1}^{2},$
$\bar{x^{3}}=3\lambda_{2}/(\beta\lambda_{1}^{2})+\lambda_{2}^{3}/\lambda_{1}^{3},$
and $\bar{x^{4}}=3/(\beta^{2}\lambda_{1}^{2})+6\lambda_{2}^{2}/(\beta\lambda_{1}^{3})+\lambda_{2}^{4}/\lambda_{1}^{4}.$
The integral over $p$ has been calculated in Eq.$\ $(\ref{seq:setequ}).
The solution of Eq.$\ $(\ref{seq:setequ}) is obtained as
\begin{eqnarray}
a_{1}(t) & = & -\frac{\dot{\lambda}_{1}}{4\lambda_{1}},\nonumber \\
a_{2}(t) & = & \frac{\dot{\lambda}_{2}}{\lambda_{1}}-\frac{3\dot{\lambda}_{1}\lambda_{2}}{4\lambda_{1}^{2}},\nonumber \\
a_{3}(t) & = & \frac{\gamma\dot{\lambda}_{1}}{8\lambda_{1}},\nonumber \\
a_{4}(t) & = & \frac{3\gamma\dot{\lambda}_{1}\lambda_{2}}{4\lambda_{1}^{2}}-\frac{\gamma\dot{\lambda}_{2}}{\lambda_{1}}.\label{seq:solutionset}
\end{eqnarray}
Then we derive the approximate auxiliary Hamiltonian in Eq.$\ $(\ref{seq:approxham})
according to Eq.$\ $(\ref{seq:appham}).

\section{The stochastic simulation\label{SAec:C}}

The dynamical evolution of the Brownian particle system is described
by the Langevin equation

\begin{eqnarray}
\dot{x} & = & \frac{\partial H}{\partial p},\nonumber \\
\dot{p} & = & -\frac{\partial H}{\partial x}-\gamma\frac{\partial H}{\partial p}+\xi(t),\label{eq:generalLaneq}
\end{eqnarray}
where $H$ is the total Hamiltonian and $\xi(t)$ is the standard
Gaussian white noise satisfying $\langle\xi(t)\rangle=0$ and $\langle\xi(t)\xi(t')\rangle=2\gamma k_{\mathrm{B}}T\delta(t-t')$.
We introduce the characteristic length $l_{\mathrm{c}}\equiv(k_{\mathrm{B}}T/\lambda_{1}(0))^{1/2}$,
the characteristic time $\tau_{1}=m/\gamma$ and $\tau_{2}=\gamma/\lambda_{1}(0)$.
Then the dimensionless coordinate, momentum, time, and control protocol
can be defined as $\tilde{x}\equiv x/l_{\mathrm{c}},$ $\tilde{p}\equiv p\tau_{2}/(ml_{\mathrm{c}})$,
$s\equiv t/\tau_{2},$ $\tilde{\lambda}_{1}\equiv\lambda_{1}/\lambda_{1}(0),$
and $\tilde{\lambda}_{2}\equiv\lambda_{2}/(\lambda_{1}(0)l_{\mathrm{c}}).$
The dimensionless Langevin equation follows as

\begin{eqnarray}
\tilde{x}' & = & \frac{1}{\alpha}\frac{\partial\tilde{H}}{\partial\tilde{p}},\nonumber \\
\tilde{p}' & = & -\frac{1}{\alpha}\frac{\partial\tilde{H}}{\partial\tilde{x}}-\frac{1}{\alpha^{2}}\frac{\partial\tilde{H}}{\partial\tilde{p}}+\sqrt{2}\zeta(s)/\alpha\text{,}\label{seq:Eq-moverdamped-rxlaeq}
\end{eqnarray}
where $\tilde{\tau}\equiv\tau/\tau_{2}$ and $\alpha\equiv\tau_{1}/\tau_{2}$.
The prime represents the derivative with respective to the dimensionless
time $s$. $\zeta(s)$ is a Gaussian white noise satisfying $\langle\zeta(s)\rangle=0$
and $\langle\zeta(s_{1})\zeta(s_{2})\rangle=\delta(s_{1}-s_{2})$.
The Euler algorithm is used to solve the Langevin equation as

\begin{eqnarray}
\tilde{x}(s+\delta s) & = & \tilde{x}(s)+\frac{1}{\alpha}\frac{\partial\tilde{H}}{\partial\tilde{p}}\delta s,\nonumber \\
\tilde{p}(s+\delta s) & = & \tilde{p}(s)-\frac{1}{\alpha}\frac{\partial\tilde{H}}{\partial\tilde{x}}\delta s-\frac{1}{\alpha^{2}}\frac{\partial\tilde{H}}{\partial\tilde{p}}\delta s+\sqrt{2\delta s}\theta(s)/\alpha\text{,}\label{seq:Eq-muverdamped}
\end{eqnarray}
where $\delta s$ is the time step and $\theta(s)$ is a random number
following the Gaussian distribution with zero mean and unit variance.
The work of the stochastic trajectories follows as

\begin{eqnarray}
\tilde{w}\equiv\frac{w}{k_{B}T} & = & \int_{0}^{1}\frac{\partial\tilde{H}}{\partial s}ds\approx\sum\frac{\partial\tilde{H}}{\partial s}\delta s.\label{seq:Eq-moverdamped-trw}
\end{eqnarray}
In simulation, we set the boundary conditions as $\vec{\lambda}(0)=(1,1)$
and $\vec{\lambda}(\tau)=(16,2).$ The parameters are chosen as $k_{\mathrm{B}}T=1$,
$\gamma=1$, and $m=0.01$. The mean work is the ensemble average
over the work of $10^{5}$ stochastic trajectories.

\end{widetext}
\end{document}